\documentclass[prb,twocolumn,showpacs,longbibliography]{revtex4-1}
\usepackage[dvips]{graphicx}
\usepackage{epsfig}
\usepackage{amsmath,amsfonts,amssymb,bm}
\UseRawInputEncoding
\setcitestyle{numbers,square}
\begin{document}
\title{Red shift of the superconductivity cavity resonance in Josephson junction qubits as a direct signature of TLS population inversion.}
\author{Alexander L. Burin} 
\affiliation{Department of Chemistry, Tulane University, New Orleans, LA 70118, USA}
\author{Moshe Schechter}
\affiliation{Department of Physics and Astronomy, Ben Gurion University of Negev,
Beer-Sheva, Israel}
\author{Daniel Tennant}
\thanks{currently at  Rigetti Computing, Berkeley, California 94710, USA}
\affiliation{Lawrence Livermore National Laboratory, Livermore, CA 94550, USA}
\author{Keith G.  Ray}
\affiliation{Lawrence Livermore National Laboratory, Livermore, CA 94550, USA}
\author{Yaniv J.  Rosen}
\affiliation{Lawrence Livermore National Laboratory, Livermore, CA 94550, USA}
\date{\today}
\begin{abstract}
Quantum two-level systems (TLSs) limit the performance  of superconducting qubits and  superconducting and optomechanical resonators  breaking down the coherence and absorbing the energy of oscillations. TLS absorption can be suppressed or even switched to the gain regime by inverting TLS populations. Here we theoretically explore the regime where the full inversion  of TLS populations is attained at energies  below a pump field quantization energy  by simultaneously applying  the pump field and the time varying bias. This regime is attained changing  the bias sufficiently slowly to fully invert TLS populations when their energies cross resonance with the pump field and sufficiently fast to avoid TLS relaxation between two resonance crossing events. This population inversion  is accompanied by a significant red shift of cavity resonance due to quantum level repulsion.  The red-shift in frequency serves as a signature of the  population inversion, as its re-entrant behavior as function of bias sweep rate and of the magnitude of the pump field allows the determination of the TLSs dipole moment and relaxation time. The predicted behavior is qualitatively consistent with the recent experimental observations in Al superconducting resonators. 
\end{abstract}

\maketitle

\section{Introduction}
\label{sec:Intr}

\begin{figure}[h!]
\centering
\includegraphics[width=\columnwidth]{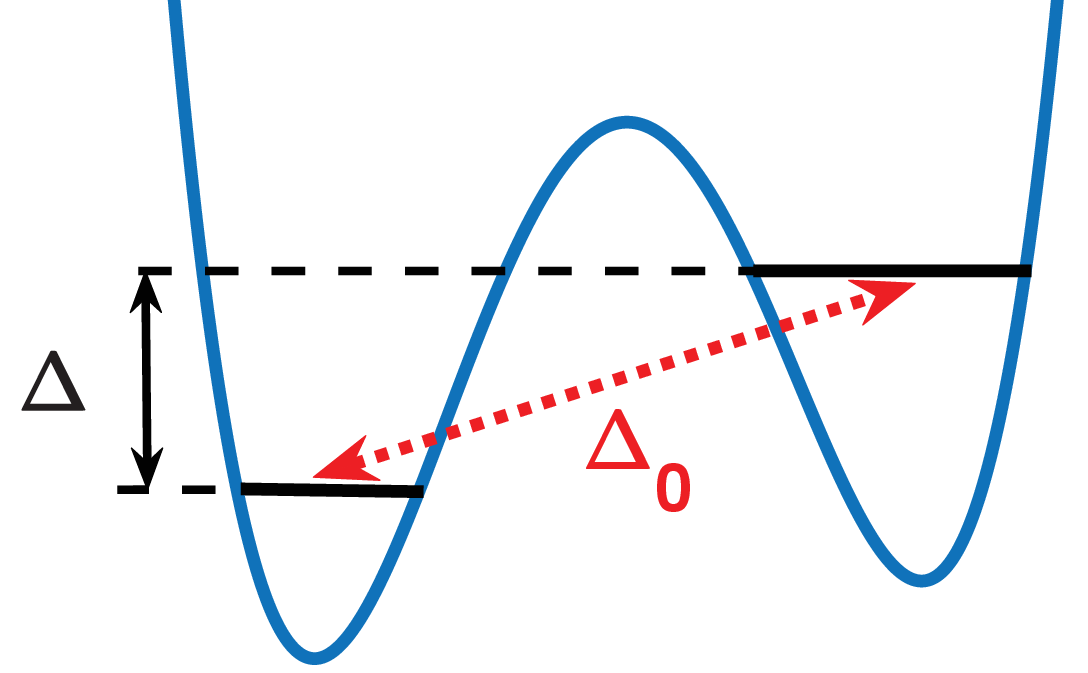}
\caption{A tunneling two-level system characterized by an asymmetry $\Delta$ and a tunneling amplitude $\Delta_0$ \cite{AHV,Ph}, determining the TLS energy as $E=\sqrt{\Delta^2+\Delta_{0}^2}$.}
\label{fig:Fig1TLS}
\end{figure}

Quantum two level systems   (TLSs) dramatically   limit coherence of quantum devices \cite{Martinis05,Ustinov2012ScienceTLS,Yu2004,Lisenfeld2019ReviewTLSs,
McDermott2009ReviewSuperQub,
oliver_welander_2013RevQub,
KlimovMartinis2018FluctEnergRelax,Bilmes2021TLSNat}.   Since the discovery of anomalous thermodynamic and kinetic properties in a variety of glasses \cite{ZellerPohl71DiscovTLS}, and their subsequent interpretation within the phenomenological Standard Tunneling Model (STM) \cite{AHV,Ph},  much effort has gone into further understanding their apparent ubiquitous nature in amorphous solids \cite{PhillipsReview,Hunklinger1986265Rev,Enss02Rev,Lisenfeld2019ReviewTLSs}. 
According to STM,   anomalous low temperature  properties of amorphous solids are determined by TLSs,  commonly believed to be  atoms or groups of atoms tunneling between two states (see Fig.   \ref{fig:Fig1TLS}).  
The STM   model is quite successful in interpretation of numerous experiments \cite{Hunklinger1986265Rev,Enss02Rev}.  In spite of  extensive  theoretical efforts  to develop the microscopic model of TLSs   \cite{Karpov83SoftPot,ParshinSoftPtoModInt94,YuLeggett88,ab96UniversJETP,ab98book,
LubchenkoWolynes2007,
Schechter2013twotls,LukinTLSPolaron,Yu2020Univers,ab23TwoTLS} the problem of TLS nature remains a long standing, yet unresolved fundamental challenge. 

TLSs are believed to exist in the Josephson junctions of superconducting qubits as well as the surface and barrier layers of superconducting and optomechanical resonators \cite{Martinis05,Cao2007NoiseInSCRes,Park2009OptMech}.  They substantially limit the maximum time of performance of those devices by means of absorbing the energy of oscillating electric field and changing chaotically the cavity resonance, thus breaking down the coherence of oscillations. The existence of TLSs in practically all disordered materials and their quantitative universality  \cite{YuLeggett88}, makes their destructive effect seemingly unavoidable.

However, one can control TLS destructive behavior even without getting rid of them.  For instance, TLS dielectric losses, relevant for superconducting qubit performance, can be controlled by means of a simultaneous application of an AC field with a frequency $\omega_{0}$, close to a cavity resonance,  and a  time varying bias. In the absence of a bias sweep,  the TLS  loss tangent  has  a maximum at small field amplitude and decreases   at larger fields, where the TLS Rabi frequency $\Omega_{R}$ exceeds its relaxation rate $T_{1}^{-1}$  \cite{VONSCHICKFUS1977144,PhillipsReview,
Martinis05,ab13LZTh}.

\begin{figure}[h!]
\centering
\includegraphics[width=\columnwidth]{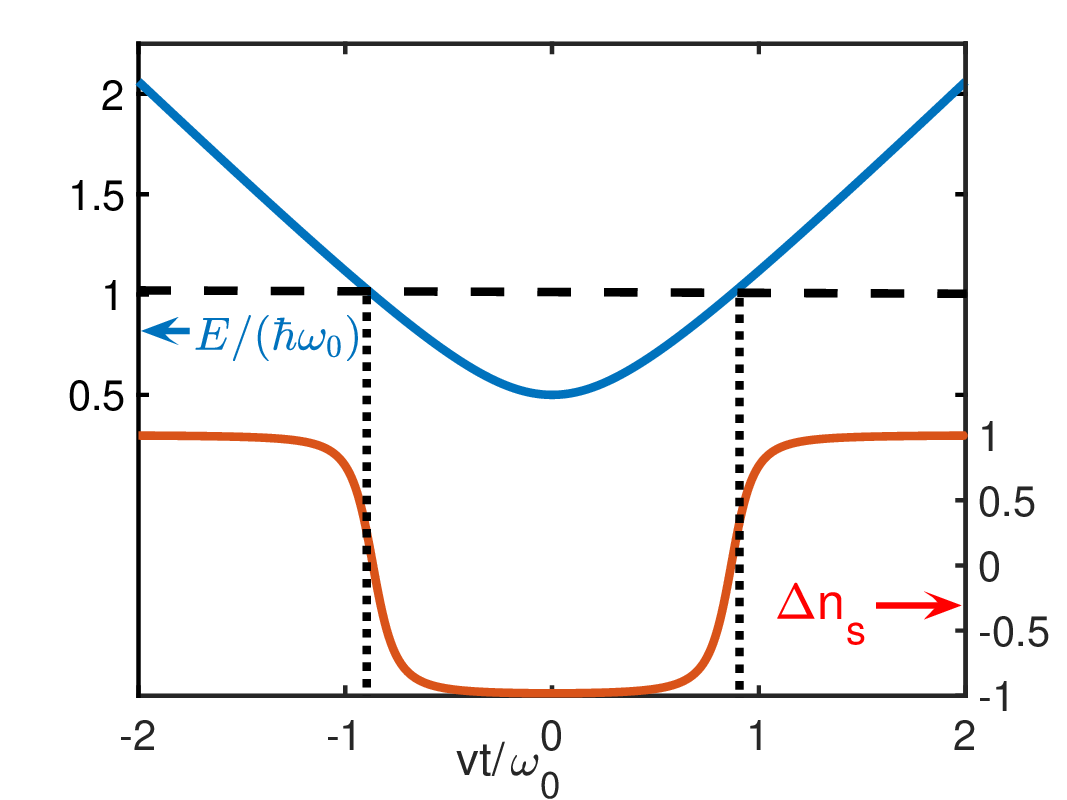}
\caption{Population inversion  of a sample TLS with $\Delta_{0}=\hbar\omega_{0}/2$ and $\Delta=\hbar vt$ occurring between adiabatic  crossings of the resonance  at $t=\mp \sqrt{E_{0}^2-\Delta_{0}^2}/(\hbar v)$ indicated by dashed lines  ($E_{0}=\hbar\omega_{0}$) in  the adiabatic Landau - Zener regime $v< \Omega_{R}^2$ and slow relaxation $vT_{1} \gg  \omega_{0}$. The top line (blue, color online) shows time dependence of TLS energy, while the bottom line (red, color online) shows time dependence of a sample TLS population  difference $\Delta n_{s}$.}
\label{fig:PopInvBas}
\end{figure} 

The application of a time varying bias, sweeping  TLS energy with a rate $v=(dE/dt)/\hbar$ (where $\hbar$ is the Planck constant),   leads to the increase  of the TLS loss tangent in the non-adiabatic Landau - Zener regime,  where this rate exceeds the square of the TLS Rabi frequency,  $\Omega_{R}^2 <v$. In this regime the loss tangent  approaches its maximum value equal corresponding to   the small AC field case of  $\Omega_{R} < 1/T_{1}$ \cite{ab13LZTh}.  In the adiabatic Landau-Zener regime the TLS population inversion takes  place  at TLS energies below resonant energy $E_{0}=\hbar\omega_{0}$  \cite{ab14LaserTheory},  i.e. the average population difference of ground and excited state $\Delta n$ gets negative. This population inversion changes loss to  gain in a certain frequency domain that can result in  lasing instability reported  in Ref. \cite{ab16TLSlaserExp}, where it was attained using a simultaneous application of two AC tones with close frequencies above and below the cavity resonance. 

Here we propose to attain the novel regime of inverted TLS populations at all energies smaller than the pump field quantization energy.  This regime is reached using a single pump field in the regime of adiabatic Landau-Zener resonance crossing $v<\Omega_{R}^2$,  if no relaxation occurs between two resonance crossings, occurring at TLS asymmetries $\Delta =\mp \sqrt{E_0^2-\Delta_{0}^2}$,   as illustrated in Fig. \ref{fig:PopInvBas}.  In this regime the  coherent oscillations. at any frequency below the pump field frequency will be not absorbed but enhanced by TLSs. due to their inverted populations. Lasing regime can is attained for resonances located below pump frequency. 
Particularly,  if the cavity possesses the acoustic resonance then the sound lasing emerges with microwave single wave pumping. 

In this novel regime the cavity resonance should be substantially red shifted, because the  population inversion pushes  downwards the cavity resonance due to quantum mechanical level repulsion as opposed to the standard regime with no population inversion. An  inverted TLS with $E<E_{0}$ pushes the resonance in an opposite direction to that for the same TLS being in the ground state \cite{Hunklinger1986265Rev}.  Thus a red shift in the cavity resonance frequency presents a direct signature of the population inversion and in the present work we investigate its dependence on the bias sweep rate and AC field amplitude. 

We found non-monotonic dependence of the frequency shift on AC field and bias field sweep rate.   The frequency first decreases with increasing the AC field and then increases back towards  its zero field value in agreement with the recent observations  in Aluminum based resonator \cite{YanivFreq}.   The theory predicts the position of the frequency minimum that   allows the determination of the typical TLS relaxation times and dipole moments based on the experimentally observed dependence.   Thus our work presents a new tool for the study of TLSs in quantum devices, and new insights into the effect of TLSs on quantum devices in and out of equilibrium conditions.

The paper is organized as follows.  The TLS model is formulated  in Sec. \ref{sec:Model}.  The expression for the frequency shift is derived and the frequency shift is calculated in Sec. \ref{sec:FrShft}, assuming the absence of TLS relaxation. Effects of TLS relaxation, lasing and field inhomogeneity are  discussed in Sec. \ref{sec:Disc}. In the conclusion Sec. \ref{sec:Concl} the results of the present paper are briefly summarized. 


 \section{Model and corrections to the resonant frequency} 
 \label{sec:Model}  

\subsection{Single TLS parameters}

Two-level systems, as represented in   Fig. \ref{fig:Fig1TLS}, enumerated by the letter $i$, can be characterized by their tunneling amplitudes $\Delta_{0i}$, asymmetry energies $\Delta_{i}$, energy splitting  $E_{i}=\sqrt{\Delta_{i}^{2}+\Delta_{0i}^{2}}$, dipole moment $\mathbf{p}_{i}$ and relaxation and decoherence times $T_{1i}$, $T_{2i}$. At low temperature,  $T\sim 10$ mK, considered in the present paper and corresponding to the performance regime of Josephson junction qubits, one has \cite{ab13LZTh}  
$T_{2}=2T_{1}$. 

In this work we consider the frequency shift within the standard tunneling model  ignoring TLS-TLS interactions which are usually believed to be weak.  In the presence of a time-varying bias this interaction modifies  the TLS loss tangent and resonant frequency due to the increase in their density of states caused by the dipole gap breakdown \cite{Osheroff94,ab95DipGap,ab98book}. The dipole gap breakdown can substantially modify TLS density of states and, consequently, TLS dielectric responses, if TLSs split into two types of weakly and strongly coupled TLSs \cite{Schechter2013twotls,ab23TwoTLS}.  Two types of TLSs were discovered in amorphous silicon \cite{ab22SiExtraTLS} and the observed loss tangent behavior in aluminum oxide  \cite{YanivFreq} have certain similarities with those observations. The wide distribution of TLS by dipole moments in aluminum oxide has been recently discovered in Ref. \cite{ab16Bahman,KDO2023NonlinTdep}. Therefore the future accurate interpretation of experimental data will possibly need an accurate consideration of TLS interaction using a dipolar gap theory \cite{ab95DipGap} and/or two TLS model \cite{Schechter2013twotls}. We ignore TLS interactions in the present work since the frequency behavior is not directly related to it. 


%

Due to quantum mechanical level repulsion, the cavity resonant frequency is sensitive to the surrounding TLS environment.  Most significant repulsion is originated from TLSs  with energies and tunneling amplitudes close to the cavity resonance $E_{i} \sim \Delta_{0i} \sim E_{0}=h\nu_{0} =\hbar\omega_{0}$.  The relaxation and decoherence times  for those TLSs are  around few microseconds  \cite{ab13LZTh,ab14LZExp}. 
 
For the temperatures under consideration the thermal energy is much smaller than the resonant energy 
\begin{eqnarray}
k_{B}T \ll \hbar\omega_{0},
\label{eq:LowT}
\end{eqnarray}
suggesting that in equilibrium all relevant TLSs are in their ground states.

\subsection{Interaction with external electric fields}

Here we consider  TLSs coupling to the parallel external AC and time varying bias electric fields, determined by TLS dipole moments.  These fields are, indeed,  nearly parallel to each other in the  experiments  \cite{ab13LZTh,ab14LZExp,YanivFreq}. 
Following Ref. \cite{ab13LZTh} we assume all TLS dipole moments having equal absolute values $p_{i}=p_{0}$  and being oriented randomly in space. The TLS dipole moment value can vary depending on the specific material. In ordinary glasses the TLS dipole moment is of order of few Debye.  The direct measurements of individual TLSs in Al oxide \cite{ab16Bahman,KDO2022ProbingIndTLSs} show a wide distribution of dipole moments from $1$ to $16$ D. The investigation of a non-equilibrium dielectric response in amorphous silicon \cite{ab22SiExtraTLS} demonstrate the existence of two types of TLSs possessing dipole moments different by orders of magnitude. with the maximum dipole moment exceeding $100$ D.  Large TLS dipole moments possibly existing in the present sample \cite{YanivFreq} can be due to the electronic nature of TLS there or tunneling clusters  \cite{LubchenkoWolynes2007}.  Here and for the rest of the paper we assume that the typical dipole moment is sufficiently large, so that the bias induced shift of TLS energy is much greater than the resonant energy $E_{0}$.  For systems with distributed dipole moments, our consideration should be valid at least qualitatively and it can be extended to those systems by averaging the result with respect to the dipole moment distribution.

The time-varying bias is applied as a sequence of identical passages, and for each of them the field changes linearly with the time at  the characteristic rate $\dot{\mathbf{F}}_{DC}$. We assume that the pulse lasts for a very long time exceeding a TLS relaxation time $T_{1}$, so the correlations between different  passages can be ignored.  Otherwise the correlations between different crossings can become important \cite{SHIMSHONI199116,Matityahu2019}. The TLS energy time dependence during a single passage  can be then  approximated  by 
\begin{eqnarray}
E_{i}(t)=\sqrt{(\Delta_{i}+\hbar v_{i}t)^2+\Delta_{i0}^2}, 
\nonumber\\
 v_{i}=v_{0}\cos(\theta_{i}), v_{0}=\frac{2\dot{F}_{DC}p_{0}}{\hbar},  
\label{eq:Et}
\end{eqnarray}
where  $\theta_{i}$ is the angle between the  TLS  dipole moment and  the $z$ axis that is parallel to the DC and AC fields. Consequently, in our calculations we assume that the bias was turned on  at time $t=-\infty$. Since the frequency shift is originated from TLSs not relaxing during a resonance passage ($vT_{1} \gg \omega_{0}$, see below, Sec. \ref{sec:FrShft}),  a typical  change of an excitation energy  of those TLSs during the time $t_{max}$ of a bias application is much greater than the resonant energy, $\hbar vt_{max} \gg E_{0}$. 

We consider an AC field characterized by its amplitude $\mathbf{F}_{AC}$ and frequency $\omega_{0}$.  Then the resonant interaction of TLSs with the AC field can be expressed  in terms of a Rabi frequencies $\Omega_{Ri}$   as  
\begin{eqnarray}
\Omega_{Ri}= \cos (\theta_{i})\frac{\Delta_{0i}}{E_{i}}\Omega_{R0}, ~
\Omega_{R0}= \frac{p_{0}F_{AC}}{\hbar},
\label{eq:OmR}
\end{eqnarray}
where  $\Omega_{R0}$  is the maximum Rabi frequency. 

We assume that electric fields are homogeneous through the sample. This approximation is especially valid in the case of using vacuum-gap capacitors of resonators \cite{KDO2023NonlinTdep}. Although this is not the case for the quoted experiment \cite{YanivFreq}, we believe that the theory should be applicable to that work at least qualitatively. 

\section{Frequency shift}
\label{sec:FrShft}

In the present work we consider only the TLS contribution to the cavity frequency ignoring the kinetic inductance associated with quasi-particles and occurring at higher fields where the excitation of quasiparticles is substantial \cite{Gaothesis,deVisser2014KinInd,ab17Katz}. The TLS induced change in the cavity resonant frequency can be found using either  the Kramers-Kronig relations of frequency shift and loss tangent \cite{Hunklinger1986265Rev} or quantum mechanical perturbation theory \cite{ab15TLSnoise}  
in the form 
\begin{widetext}
\begin{equation}
\frac{\delta\omega}{\omega}= -\frac{4\pi P_{0}p_{0}^2}{\epsilon}\int_{0}^{1}d\cos \theta(\cos \theta)^2{\rm PV}\int_{0}^{E_{max}}\frac{dE}{E}\int_{0}^{{\rm min}(E, E_{0})} \frac{\Delta_{0}d\Delta_{0}}{\sqrt{E^2-\Delta_{0}^2}} \frac{(\Delta n(E, \Delta_{0}, \theta)-1)}{(E-E_{0})}.
\label{eq:FrShft_neq}
\end{equation}
\end{widetext}
where $P_{0}$
is a TLS density of states,  $\Delta n(E, \Delta_{0}, \theta)$ is the population difference for any given TLS with an energy $E$, a tunneling amplitude $\Delta_{0}$ and an angle $\theta$ between TLS dipole moments and the electric fields. Unity  is subtracted from the population difference  for any given TLS to consider only the correction associated with the time-varying bias and the AC field.  Without the applied bias,  the population difference can be set equal  to unity for all relevant energies because of the low temperature Eq. (\ref{eq:LowT}). In the absence of the bias sweep, the frequency is practically not sensitive to the AC field  \cite{Hunklinger1986265Rev},  since the field dependence of the population difference  at $|E-\hbar\omega_{0}| < \Omega_{R0}$ does not affect the principal value integral (PV) over energy. The upper integration limit over $\Delta_{0}$ for TLSs with energies exceeding the resonant energy $E_{0}$ is set to the resonant energy $E_{0}$ to consider only TLSs undergoing resonance crossing.  

In the absence of relaxation one can approximate a TLS population difference using Landau-Zener theory  for the crossing of resonance at $E=E_{0}$. For any TLS with an energy $E <E_{0}$, a tunneling amplitude $\Delta_{0}$ and an angle $\theta$ between its dipole moment $p_{0}$ and the bias field, its population difference is determined by the level crossing at $E(t)=E_{0}$. This crossing can be characterized by the energy sweep rate $\hbar |v|=|dE/dt| = \hbar v_{0}|\cos(\theta)|E_{0}/\sqrt{E_{0}^2-\Delta_{0}^2}$ and the coupling strength $\hbar\Omega_{R}=\hbar\Omega_{R0}\cos(\theta)\Delta_{0}/E_{0}$ cf. Eqs. (\ref{eq:Et}), (\ref{eq:OmR}) and Ref. \cite{ab13LZTh}.  Then the probability that after the resonance crossing TLS remains in the same state  (ground or excited) is given by 
\begin{eqnarray}
P_{LZ}=e^{-\frac{\pi \Omega_{R}^2}{2v}}. 
\label{eq:LZPr}
\end{eqnarray}
Consequently, the TLS population difference for energies below the resonant energy can be expressed as 
\begin{eqnarray}
\Delta n = 2P_{LZ}-1, ~~ E<E_{0}. 
\label{eq:LZPopDiffIns}
\end{eqnarray}

TLSs possessing energies exceeding the resonant energy ($E>E_{0}$) approach this state either without level crossings or with two subsequent Landau-Zener level crossings (the possibility of a single level crossing can be neglected since  the typical change of TLS  excitation energies due to a time varying  bias  is much greater than the resonant energy $E_{0}$, see Sec. \ref{sec:Model}). In the former case their population difference is approximately $1$, while in the latter case it is given by $P_{LZ}^2+(1-P_{LZ})^2-2P_{LZ}(1-P_{LZ})=(1-2P_{LZ})^2$. The average population difference can then be expressed as 
\begin{eqnarray}
\Delta n = ((1-2P_{LZ})^2+1)/2, ~~ E>E_{0}. 
\label{eq:LZPopDiffOut}
\end{eqnarray}

Landau-Zener expressions for the population differences are not applicable in the vicinity of the crossing point $|E-E_{0}| \leq \hbar \Omega_{R}$ for $v < \Omega_{R}^2$ or $\sqrt{v}$ otherwise,  where the population difference changes continuously. These inequalities determine the lower cutoff for the logarithmic integral over energy in Eq. (\ref{eq:FrShft_neq}) at $|E-E_{0}| \sim {\rm max}(\hbar\Omega_{R}, \hbar\sqrt{v})$ while the upper cutoff is given by $|E-E_{0}| \approx E_{0}$.  Within the logarithmic accuracy one can replace ${\rm max}(\Omega_{R}, \sqrt{v}) \approx \Omega_{R}+\sqrt{v}$. Then the integrals over energy in Eq. (\ref{eq:FrShft_neq}) can be evaluated with the logarithmic accuracy using the definitions of the population differences Eqs. (\ref{eq:LZPopDiffIns}) and (\ref{eq:LZPopDiffOut}). Finally, we arrive at
\begin{widetext}
\begin{equation}
\frac{\delta\omega}{\omega}= -\chi\int_{0}^{1}d\cos(\theta)\cos(\theta)^2\int_{0}^{E_{0}} \frac{\Delta_{0}d\Delta_{0}}{E_{0}\sqrt{E_{0}^2-\Delta_{0}^2}} \left(1-e^{-\frac{\pi |\cos(\theta)|)\Delta_{0}^2\Omega_{R0}^2}{2E_{0}\sqrt{E_{0}^2-\Delta_{0}^2}v_{0}}}\right)^2\ln\left(\frac{\omega_{0}}{\Omega_{R0}+\sqrt{v_{0}}}\right), ~ \chi=\frac{8\pi P_{0}p_{0}^2}{\epsilon}.
\label{eq:FrShft_neqLogAccr}
\end{equation}
\end{widetext}

The integral in Eq. (\ref{eq:FrShft_neqLogAccr}) can be evaluated analytically in two limiting regimes of a small and large Landau-Zener parameter $\xi=\pi \Omega_{R0}^2/(2v_{0})$ compared to unity. These two regimes correspond to adiabatic and non-adiabatic level crossings respectively. In those two limits the results can be expressed as 
\begin{eqnarray}
\frac{\delta\omega}{\omega}\approx -\chi\ln\left(\frac{\omega_{0}}{\Omega_{R0}+\sqrt{v_{0}}}\right)
\begin{cases}
\frac{0.3466\pi}{2} \frac{\Omega_{R0}^2}{2v_{0}}, ~ 
\Omega_{R0}^2 \ll v_{0}, \\
\frac{1}{3},  ~\Omega_{R0}^2 \gg v_{0}.
\end{cases}
\label{eq:asympt}
\end{eqnarray}
Thus the absolute value of the negative correction to the frequency increases with the AC field amplitude (number of photons in the cavity) at small fields proportionally to the probability to excite TLSs passing through resonance, which is  proportional  to $\Omega_{R0}^2.$  
At large fields $\Omega_{R0}^2 \gg v_{0}$ the correction to the frequency decreases logarithmically with that field since the occupation across the resonance changes continuously over an energy proportional to $\Omega_{R0}$. The minimum of frequency, or maximum correction, emerges at $\Omega_{R0}^2 \approx v_{0}$.  

\begin{figure}[h!]
\centering
\includegraphics[width=\columnwidth]{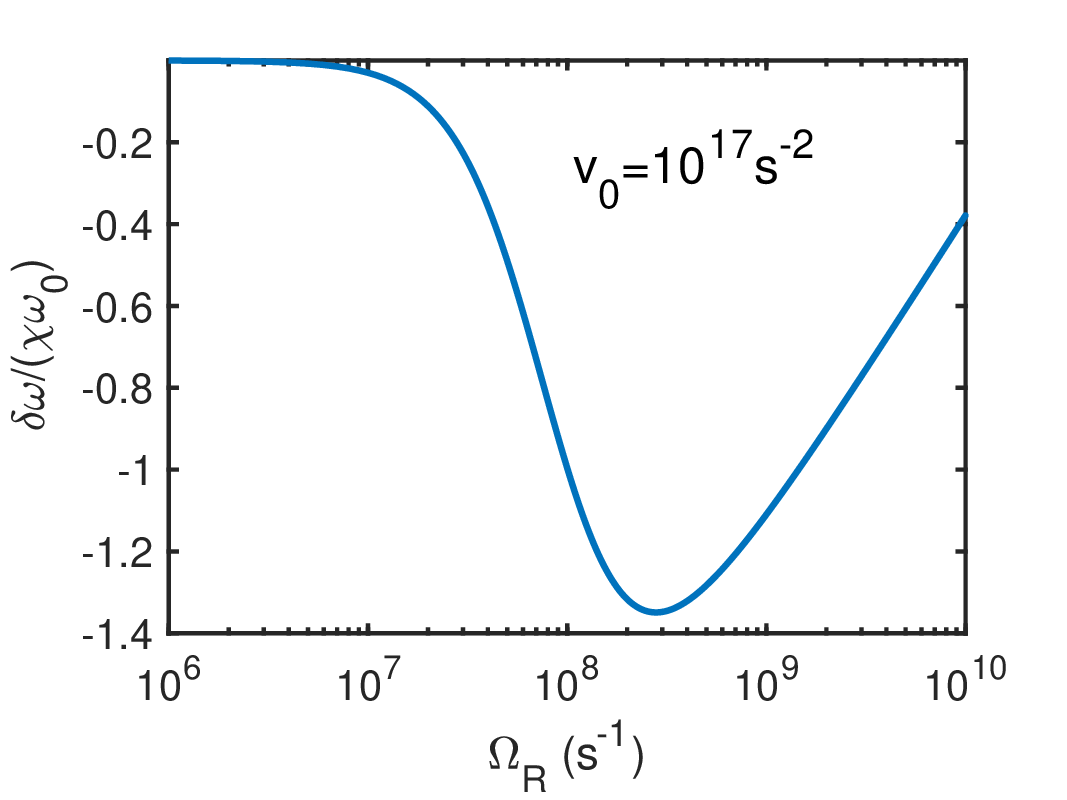}
\caption{Frequency shift vs. Rabi frequency. The dimensionless factor $\chi =8\pi P_{0}p^2/\epsilon$ (cf. Eq. (\ref{eq:FrShft_neqLogAccr})) is of order of $10^{-4}$.}
\label{fig:FrOmR}
\end{figure}

\begin{figure}[h!]
\centering
\includegraphics[width=\columnwidth]{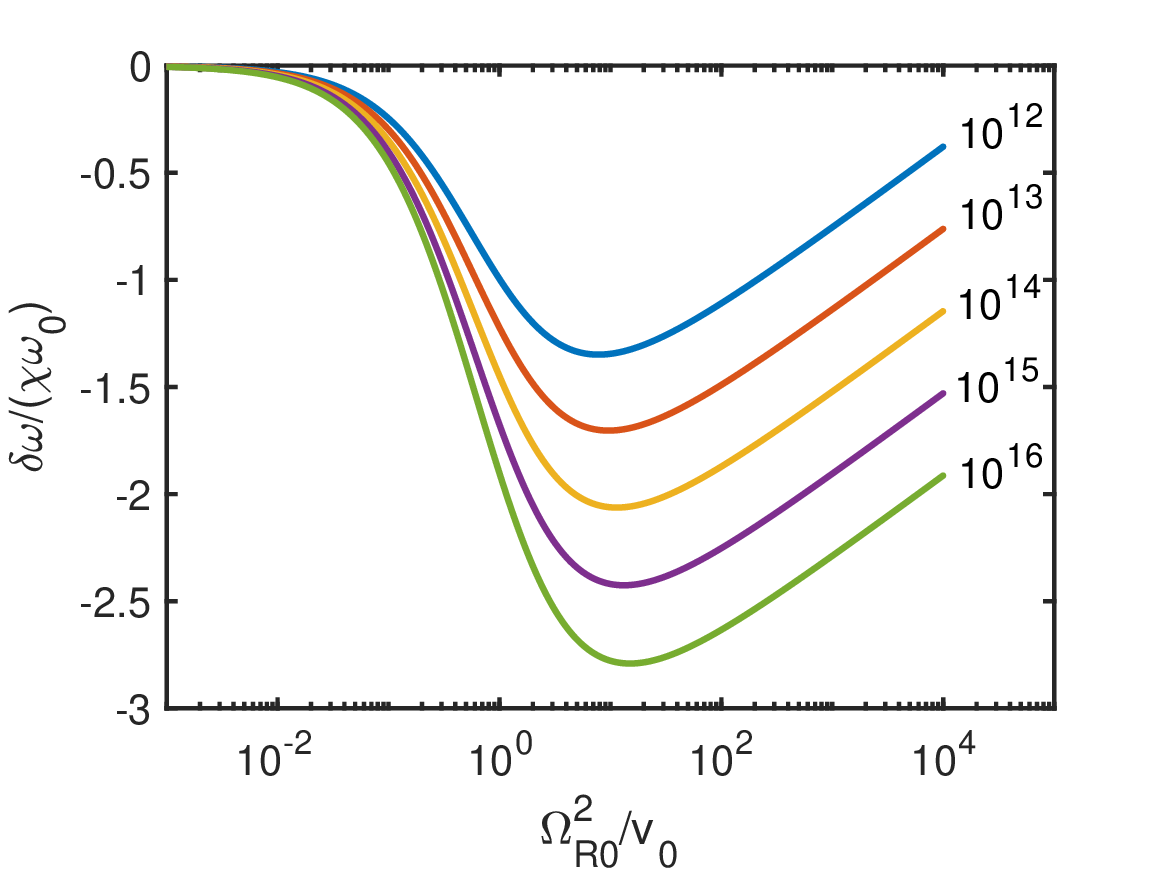}
\caption{Frequency shift vs.  an effective Landau-Zener parameter ($\Omega_{R0}^2/v_{0}$)  for different bias sweep rates given on the right hand side of each graph in s$^{-2}$.}
\label{fig:FrOmRmany}
\end{figure}

\begin{figure}[h!]
\centering
\includegraphics[width=\columnwidth]{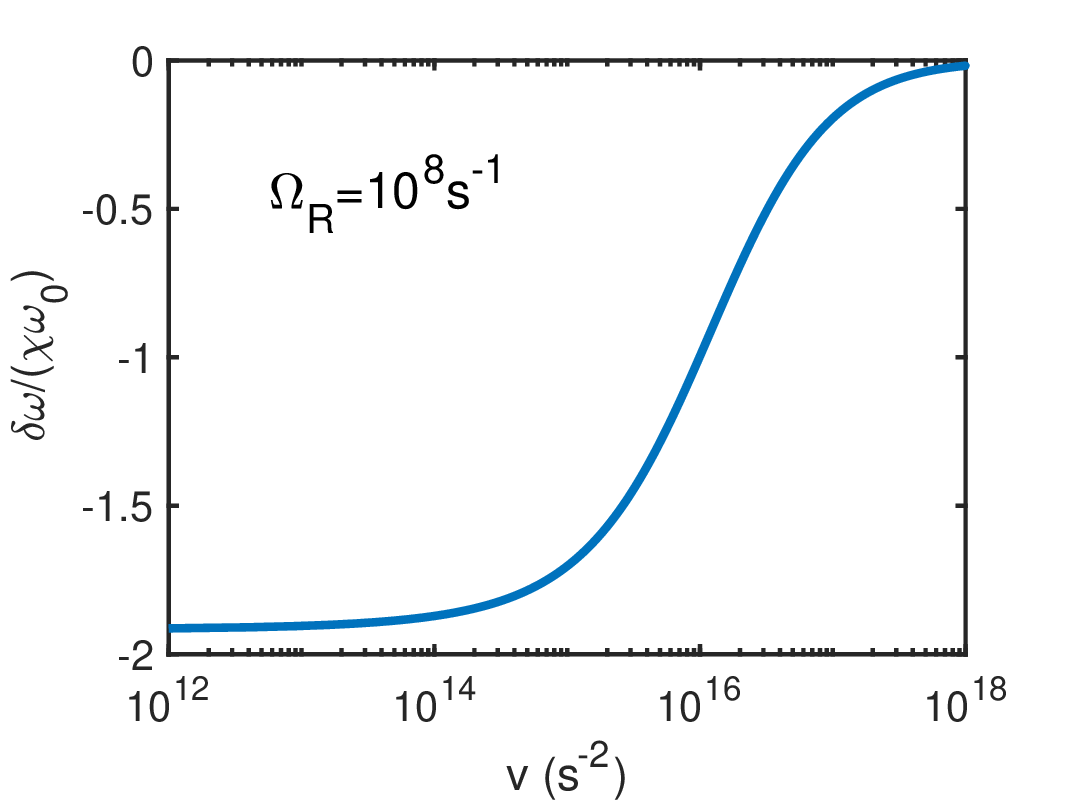}
\caption{Frequency shift vs. bias sweep rate.}
\label{fig:Frv}
\end{figure}

The dependencies of the frequency shift on a Rabi frequency and a  bias sweep rate obtained integrating numerically Eq. (\ref{eq:FrShft_neqLogAccr}) are presented in Figs. \ref{fig:FrOmR}, \ref{fig:FrOmRmany} and \ref{fig:Frv}. For the illustration of a typical Rabi frequency dependence of the frequency shift, Fig. \ref{fig:FrOmR},  we choose the  bias sweep rate given by $v_{0}=10^{17}$s$^{-2}$. For this choice the typical TLS energy change during its relaxation time $\delta E \sim \hbar v_{0}T_{1}$ ($T_{1}\approx 3\cdot 10^{-6}$s  \cite{ab14LZExp}) exceeds the TLS resonant energy $E_{0}=\hbar\omega_{0}$ ($\omega_{0} \sim 2\pi\cdot 5\cdot 10^{9}$s$^{-1}$ \cite{ab14LZExp}) by an order of magnitude, so the TLS relaxation during the resonance domain passage time $\omega_{0}/v$ can be neglected. The maximum frequency shift (minimum of frequency) is found  at $\Omega_{R} \approx 2\sqrt{v_{0}}$. Since the bias sweep rate  Eq. (\ref{eq:Et}) and the Rabi frequency Eq. (\ref{eq:OmR}) are determined by known AC field amplitude $F_{AC}$ and bias field sweep rate $\dot{F}_{DC}$ one can estimate a typical TLS dipole moment as 
\begin{eqnarray}
p=8\frac{\hbar\dot{F}_{DC}}{F_{AC*}^2}, 
\label{eq:DipMomEst}
\end{eqnarray} 
where $F_{AC*}$ is the AC field amplitude  corresponding to the maximum resonant frequency shift.

This estimate is weakly sensitive to the choice of the bias sweep rate as is illustrated in Fig. \ref{fig:FrOmRmany}, where the frequency shift  is given  as a function of the effective Landau Zener parameter ($\Omega_{R0}^2/v_{0}$) similarly to Refs. \cite{ab13LZTh,ab14LZExp}, where a similar rescaling is used for the loss tangent. However,  the frequency shift dependencies  don't collapse to the same curve  in contrast to that for the loss tangent  due to the non-universal logarithmic factors lacking for the loss tangent.   Still the dimensionless ratio of the Rabi frequency and the square root of the bias sweep rate $\Omega_{R0}/\sqrt{v_{0}}$, corresponding to the scaled value of the maximum frequency shift, changes by  25\% for a bias sweep rate change by six orders of magnitude. 

We note that the dependencies observed experimentally can be less universal due to the relaxation destroying the population inversion as discussed below in Sec. \ref{sec:Relax}.  

The frequency shift decreases constantly with increasing bias sweep rate, as illustrated in Fig. \ref{fig:Frv} for the Rabi frequency $\Omega_{R}=10^{8}$s$^{-1}$. This is due to the weakening of the population inversion for
 larger bias sweep rates. A monotonic increase of the frequency shift with decreasing the bias sweep rate will be relevant until sufficiently  small bias sweep rates $v_{0}< \omega_{0}/T_{1}$, where relaxation gets significant as discussed below in Sec. \ref{sec:Relax} (see Fig. \ref{fig:FrvRel}). 

\section{Discussion of effects left out, but significant under certain conditions.}
\label{sec:Disc}

\subsection{Frequency shift in the presence of TLS relaxation}
\label{sec:Relax}

At small bias sweep rate $v_{0} < \omega_{0}/T_{1}$,  TLSs transfer to their ground state between two passages of resonances. Then the contributions to the frequency shifts of population inversions during  two successive passages substantially compensate each other.  Only TLSs not relaxing to the ground state during the passage between two resonances contribute to the frequency shift. Thus each TLS contribution to the frequency shift in Eq. (\ref{eq:FrShft_neq}) should be weighted by the probability that no relaxation takes place during the passage. This probability can be expressed in terms of energy and tunneling amplitude dependent relaxation time $T_{1}(E, \Delta_{0})$ as 
\begin{eqnarray}
P=\exp\left[-\int_{t_{1}}^{t_{2}}\frac{dt}{T_{1}(E(t), \Delta_{0})}\right], 
\label{eq:RelProb}
\end{eqnarray} 
where first and second resonance crossings, $E(t)=E_{0}$, take place at times $t_{1}$ and $t_{2}$, respectively (see Eq. (\ref{eq:Et})).

 For the temperature domain of interest given in Eq. (\ref{eq:LowT}), one can approximate a TLS relaxation time dependence on energy and tunneling amplitude as given in Refs. \cite{Jackle1972,ab13echo} 
\begin{eqnarray}
T_{1}(E, \Delta_{0})=T_{10}\frac{E_{0}^3}{E\Delta_{0}^2}, 
\label{eq:T1}
\end{eqnarray} 
where $T_{10} \sim 3\mu$s \cite{ab14LZExp}  is the relaxation time of TLS with energy and tunneling amplitude equal to the resonant energy.  Then the integration in Eq. (\ref{eq:RelProb}) can be performed analytically and we get 
\begin{widetext}
\begin{eqnarray}
P(E, \Delta_{0}, v_0, \cos(\theta))=
\exp\left[-\frac{\Delta_{0}^2}{E_{0}^2}\frac{\omega_{0}}{v_{0}T_{10}|\cos(\theta)|}\left(\frac{\Delta_{0}^2}{E_{0}^2}\ln\left(\sqrt{\frac{E_{0}^2}{\Delta_{0}^2}-1}+\frac{E_{0}}{\Delta_{0}}\right)+\sqrt{1-\frac{\Delta_{0}^2}{E_{0}^2}}\right)\right].
\label{eq:RelaxAnal}
\end{eqnarray} 
\end{widetext}
 This exponent should be added to the integrand in the definition of the frequency shift Eq. (\ref{eq:FrShft_neq}). 

\begin{figure}[h!]
\centering
\includegraphics[width=\columnwidth]{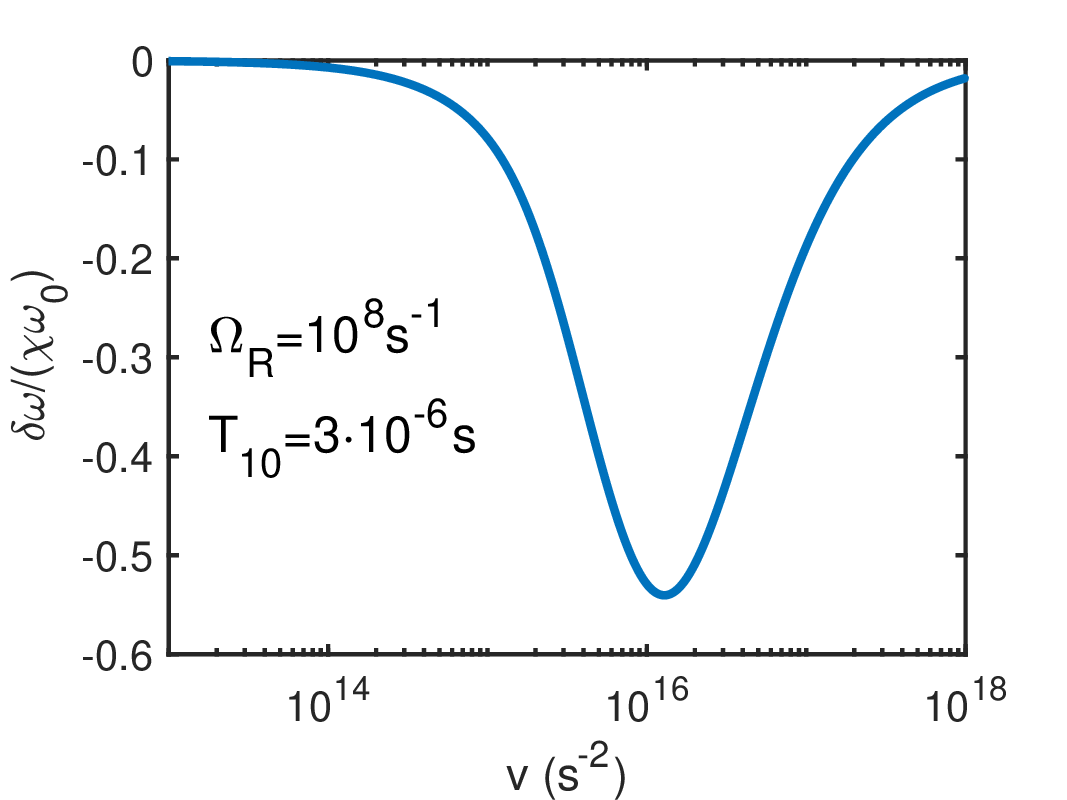}
\caption{Frequency shift vs. bias sweep rate in the presence of relaxation.}
\label{fig:FrvRel}
\end{figure} 


The logarithmic integral can be evaluated as in the absence of relaxation.  However, for $\omega_{0}>v_{0}T_{10}$ the exponent in Eq. (\ref{eq:RelaxAnal}) suppresses the contributions of  TLSs with tunneling amplitudes exceeding $\Delta_{01}=E_{0}\sqrt{\omega_{0}/(vT_{10})}$. Consequently, the argument of the logarithm in Eq. (\ref{eq:FrShft_neqLogAccr}) should be modified replacing the maximum Rabi frequency with the smaller value $\Omega_{R0}/\sqrt{\omega_{0}/(vT_{10})}$. The common replacement $\Omega_{R0} \rightarrow \Omega_{R0}/\sqrt{1+\omega_{0}/(vT_{10})}$ describes all regimes within the logarithmic accuracy. In Fig. \ref{fig:FrvRel}, the frequency dependence on the bias sweep rate is shown for the largest considered  AC field  ($\Omega_{R0}=10^{8}$s$^{-1}$) and resonant TLS relaxation time $T_{10} =3\cdot 10^{-6}\mu$s \cite{ab14LZExp}. As  expected at small bias sweep rates the frequency shift disappears due to TLS relaxation. Therefore this dependence has a minimum at the bias sweep rate $v_{0} \approx \omega_{0}/T_{10}$ in contrast to the monotonic dependence in the absence of relaxation (see Fig. \ref{fig:Frv}). The position of the minimum can be used to determine a typical TLS relaxation time $T_{10}$.

Similarly to Eq. (\ref{eq:asympt}) we define two asymptotic behaviors  of the cavity frequency in the non-adiabatic and adiabatic Landau-Zener regimes in the presence of relaxation as 
\begin{widetext}
\begin{eqnarray}
\frac{\delta\omega}{\omega}\approx -\chi\ln\left(\frac{\omega_{0}}{\frac{\Omega_{R0}}{\sqrt{1+\omega_{0}/(vT_{10})}}+\sqrt{v_{0}}}\right)
\begin{cases}
\frac{\pi^2}{32} \frac{\Omega_{R0}^4}{v_{0}^2}\frac{v_{0}^3T_{10}^3}{\omega_{0}^3}, ~ 
 \frac{v_{0}T_{10}}{\omega_{0}}\Omega_{R0}^2  \ll v_{0}, \\
\frac{v_{0}T_{10}}{8\omega_{0}},  ~~~~~~~~~~ \frac{v_{0}T_{10}}{\omega_{0}} \Omega_{R0}^2 \gg v_{0}.
\end{cases}
\label{eq:asympt1}
\end{eqnarray}
\end{widetext}
In the presence of relaxation, frequency also approaches minimum as a function of Rabi frequency at $\frac{v_{0}T_{10}}{\omega_{0}}\Omega_{R0}^2  \approx v_{0}$. This minimum is smaller by the absolute value  compared to that  in the absence of relaxation by the factor of $v_{0}T_{10}/\omega_{0}$. It is shifted towards larger Rabi frequencies. The main contribution to the frequency shift is due to TLSs with tunneling amplitudes $\Delta_{0}\leq \Delta_{01} \approx E_{0}\sqrt{v_{0}T_{10}/\omega_{0}}$, as TLSs with larger tunneling amplitude transfer to the ground state during the passage between resonances. 


\subsection{Lasing} 
\label{sec:Las}

For fast bias  sweeps, $v_{0} > \omega_{0}/T_{1}$, yet  still in the adiabatic Landau-Zener regime $\Omega_{R0}^2 \gg v_{0}$ one obtains  population inversion of TLSs in the energy domain $E< E_{0}$ (see Fig. \ref{fig:PopInvBas}). If the cavity resonance belongs to the energy domain  of the inverted  population, this can lead to a lasing instability similar to Ref. \cite{ab16TLSlaserExp,ab14LaserTheory} if the external or background losses are smaller compared to the internal ones. Then one should answer the question: how does lasing affect the frequency shift? 

It seems that the position of the cavity resonance should be insensitive to lasing. This is because lasing is impossible if the AC field frequency $\omega_{0}$ is less than the cavity resonant frequency $\omega_{c}$ as  there is no population inversion at cavity resonance. Then for $\omega_{0} < \omega_{c}$  the previous consideration is valid and the  cavity resonance should remain in the same position  as in the absence of lasing. However, at $\omega_{0} > \omega_{c}$, lasing can affect  the frequency shift Eq. (\ref{eq:FrShft_neq}) due to the gain saturation. It can significantly modify the shape of the frequency dependent system response to the AC field compared to the regime with no lasing, yet, without changing the position of the frequency minimum.  

We postpone the investigation of the system response in a lasing regime for the future.  According to Ref. \cite{YanivFreq} no lasing was observed in the experiment, where the frequency shift was discovered. This can be due to a relatively small quality factor of the cavity,  a significant field inhomogeneity briefly discussed below in Sec. \ref{sec:inhom}  or because only a fraction of the TLSs switch.   

\subsection{Field inhomogeneity}
\label{sec:inhom}

The field inhomogeneity existing in the original experiment \cite{YanivFreq} can influence the experimental observations modifying frequency dependencies compared to the present model   (see. Figs. \ref{fig:FrOmR}, \ref{fig:FrOmRmany}, \ref{fig:Frv}). The field inhomogeneity creates an effective distribution of Rabi frequencies.  
The results for the frequency shift in Eq. (\ref{eq:FrShft_neq})  and the asymptotic behaviors in  Eq. (\ref{eq:asympt}) should be averaged over that  distribution. This would result in slower reduction  of the frequency shift at small fields. Particularly, if the field probability density function inside the sample possess a long power law tail $P(F)\propto F^{-\alpha}$ with $1 <\alpha <3$ then the small field asymptotic behavior $\delta\omega \propto F_{AC}^2$ Eq. (\ref{eq:asympt}) will change to  $\delta\omega \propto F_{AC}^{3-\alpha}$ in a broad field domain preceding the minimum. This anomalous behavior is possibly relevant for the experimental work \cite{YanivFreq}. The logarithmic  dependence at large fields  should remain valid. 

\section{Conclusion}
\label{sec:Concl}

In this work the novel non-linear response of two level systems to the external resonant AC field in the presence of  fast time varying bias has been explored, where TLS population inversion takes place for TLSs possessing energies lower than the cavity resonance energy.   In this regime the absorption by TLSs in the frequency domain below the pump field frequency is replaced  with the gain supporting coherent oscillations that might significantly improve the performance of quantum resonators at those frequencies.  This regime is realized if the bias sweep is fast enough, so that the TLS energy passes the resonant domain faster than  the TLS relaxation time.  Then the cavity resonant frequency, or TLS contribution to the real part of the  dielectric constant, decreases with increasing the microwave field due to TLS population inversion. This occurs  until the  adiabatic Landau-Zener regime is approached, where the frequency reaches its minimum and then increases  with a further increase of the microwave field back to its original value.   The apparent non-monotonic red-shift of the cavity resonance thus serves as a direct signature for the occurance of TLS population inversion in this regime. Furthermore, the position of the minimum can be used to estimate a typical TLS dipole moment and/or TLS relaxation time in the regime of significant relaxation.    

The predicted behavior is qualitatively consistent with the recent experimental observations \cite{YanivFreq}. 
More accurate verification of the theory can be performed using samples with a homogeneous fields  realized in  vacuum-gap capacitors of resonators \cite{KDO2023NonlinTdep}. 

A cavity can reach lasing instability  in the adiabatic Landau - Zener regime corresponding to high AC fields  (see Fig. \ref{fig:FrOmR}).  Using high quality cavity with homogeneous fields, lasing can be realized  using a  single tone, in contrast with the previously discovered two tone microwave TLS laser \cite{ab16TLSlaserExp}.  If the cavity possesses the acoustic resonance then the system will function as a sound laser. 




\begin{acknowledgments}
The work of A. B. is supported by the National Science Foundation (CHE-2201027) and the NSF Epscore and Louisiana Board of Regents LINK program.  M. S.  acknowledges financial support from the ISF (Grant no. 2300/19).  
Work at LLNL was supported by LDRD 20-ERD-010 and was performed under the auspices of
the U.S. Department of Energy by Lawrence Livermore National Laboratory under Contract DE-AC52-07NA27344.


\end{acknowledgments}
 






\bibliography{MBL1}

\begin{thebibliography}{46}
\bibitem [{\citenamefont {Moncrieff}\ and\ \citenamefont
  {Rozen}()}]{YanivFreq}%
  \BibitemOpen
  \bibfield  {author} {\bibinfo {author} {\bibfnamefont {Daniel}\ \bibnamefont
  {Moncrieff}}\ and\ \bibinfo {author} {\bibfnamefont {Yaniv}\ \bibnamefont
  {Rozen}},\ }\href@noop {} {}\bibinfo {note} {To be published
  elsewhere}\BibitemShut {NoStop}%
\end{thebibliography}
\end{document}